\documentstyle[twoside,fleqn,espcrc2]{article}

\newcommand{\AmS}{{\protect\the\textfont2
  A\kern-.1667em\lower.5ex\hbox{M}\kern-.125emS}}

\title{Exotic hadronic states and all-to-all quark propagators}

\author{UKQCD Collaboration: Chris Michael  and  Janne Peisa   
\address{Theoretical Physics Division, 
        Department of Mathematical Sciences, \\ 
        University of Liverpool, Liverpool L69 3BX, United Kingdom}%
   \thanks{Supported by PPARC grant GR/K 53475}
}

\begin{document}

\begin{abstract}
 We discuss methods to obtain accurate hadronic spectra with propagating
quarks. Comparing the determination of masses for spin-exotic hybrid
mesons  with glueball mass determinations, we conclude that quark
propagators from all sites to all other sites would enable great
improvement in the errors. Such  propagators are achievable by using
stochastic estimators. We discuss  previous attempts and present our
method for maximal variance reduction.  This is a very promising
technique and we illustrate it by obtaining the  spectrum of ground
state and excited B mesons in the limit where the $b$ quark is static.

\end{abstract}

\maketitle

\section{INTRODUCTION}

 One of the goals of lattice QCD is to pin down the hadronic spectrum 
accurately in the quenched approximation and then to explore the shifts 
as the dynamical quark degrees of freedom are re-introduced. This is 
particularly relevant to guide experimental searches for glueballs 
and hybrid mesons. The situation for lattice determination of the 
spectrum of hybrid mesons, at present, is that the first clear signals 
have been obtained but much greater precision would be valuable.

 It is interesting to contrast this situation with the glueball mass
determination for which a precise continuum value has been
obtained~\cite{ukqcd,gf11} in the quenched approximation. The glueball
correlator is measured from all sites on the lattice which  gives large
statistics even from one gauge configuration. For correlations involving
gauge links only, this is very easy to achieve computationally. For 
correlations involving fermion propagation, the inversion of the fermion
matrix from  even a single source is computationally expensive. Thus  
the hybrid correlator has been measured~\cite{hybrid,hybus} from only a
single source for  each gauge configuration which gives a rather noisy
signal.  Moreover to  construct appropriate non-local sources for hybrid
meson studies can also involve  extra quark propagator inversions which
is costly. The same constraints also apply to a detailed study 
of the orbital and radial excited hadrons as well.

\begin{figure}[t]
\vspace{2.4in}
\includegraphics{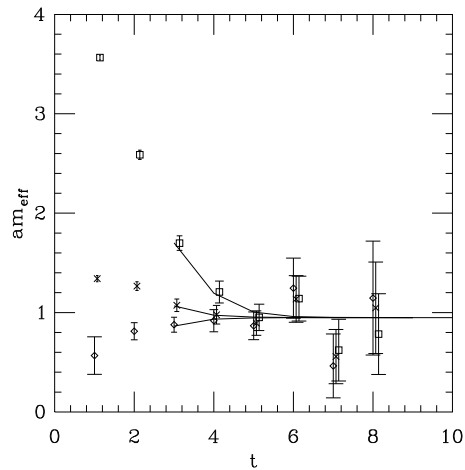}

\caption{  The  effective mass of the $J^{PC}=1^{-+}$ hybrid meson 
versus $t$ from the UKQCD data~{\protect\cite{hybrid}} with U-shaped
source and several different sinks.}

\end{figure}

 The result of the recent UKQCD determination~\cite{hybrid} of the
spin-exotic  $J^{PC}=1^{-+}$ hybrid meson is shown in Figure~1. This was
obtained from 350 propagators from each of  two sources separated by a 
fixed distance  and joined by a U-shaped gauge link. By using different
operators at the sink, several different correlators were evaluated
which allows a separation of ground state  and excited states of the
required quantum numbers. The study  with a clover-Wilson fermion action
using  $C_{SW}=1.4785$ at $\beta=6.0$ and a hopping parameter
corresponding to a quark mass close to  strange, gave a hybrid meson
mass of 2.0(0.2) GeV for the $\bar{s} s$ state. The results  for other
exotic spin-parity hybrid mesons are consistent with being heavier than
this $1^{-+}$ state.  

As already indicated, a more precise determination of these spectra
would  be valuable. With this in mind, we now discuss the feasibility of
evaluating  all-to-all hadronic correlators. As an illustration of the
potency of this  approach, we will present a preliminary study of the
spectrum of excited  $\bar{b}q$ mesons. Here spin-exotic hybrid mesons
cannot be identified because charge conjugation C is not a good quantum
number. Nevertheless, lattice  studies can guide experimental searches
in this area where few experimental results are available.

\section{STOCHASTIC PROPAGATORS}

    The conventional approach to quark propagation in a gauge background
field  (this discussion applies equally to the quenched case and to full
QCD with dynamical quarks) is to solve the lattice Dirac equation $Q
\psi=\delta$ iteratively from a single source  to determine a
propagator $G$ satisfying $Q G = 1$. This problem is equivalent to that
of the  inversion of a sparse matrix  since $Q$ has only local and
nearest neighbour contributions -- this inversion is solved by 
iteration which is computationally expensive. For a local source at 
$x$,  the propagator $G(y,x)$ contains information about the quark
propagation  from this source to all lattice sites $y$. For the
Wilson-Dirac discretisation (this includes the  SW-clover case) then
$Q(x,y)= \gamma_5 Q(y,x)^{*} \gamma_5$ and  hence the propagation from
all sites to the  source is also obtained. Thus in the usual approach,
only propagators to or from  $x=(0, 0, 0, 0)$ are determined. This
limits the correlations that can be constructed  to explore hadronic
spectra and matrix elements and can cause large statistical errors and
large correlations among different $t$ values since so few samples of
the correlation are evaluated for each  gauge configuration.

In the  case of a  meson with one static quark and one propagating light
quark, then only the elements  of $G$ with the same spatial component
$\bf{x}=\bf{y}$ will be used in  such correlations and  two such
numbers (from positive and negative time directions) per gauge
configuration  will be available for use to study the correlation versus
time $t$. Thus only a very small part of the information obtained
labouriously by inverting $Q$ to  obtain $G$ is  used in this extreme
case. An attractive alternative is to obtain the propagator from many
different sites.  The default way to achieve this is just to repeat the
inversion for sources at  different  positions. This is computationally 
demanding and, if the propagators were to be stored, would involve
prodigious storage requirements for all-to-all propagators.
 
An alternative route is to accept stochastic estimators of the
propagators.  Since the fermion matrix inversion is conducted in a gauge
field sample, it  is not necessary  to determine the propagator $G$ to
10 digit accuracy as is  usually done. Indeed, varying the elements of
the propagator randomly by a few percent makes little change to  the
hadronic correlators obtained. The problem is to achieve this stochastic
sampling in an unbiased way. Just stopping the inversion algorithm at
relative accuracy  of 1\% introduces bias. One acceptable approach is 
to use a pseudo-fermion method with scalar fields interacting via 
the fermion matrix.  This amounts to obtaining a stochastic
estimate  of the inverse of a matrix $M$ by use of the gaussian Monte
Carlo process:
 \begin{equation}
  Z=\int D \phi \ e^{-{1 \over 2}\phi^* M  \phi }
 \end{equation}
 with $  M^{-1}_{ji}= \, < \phi_i^* \phi_j  > $. This approach requires
a positive definite matrix $M$ which  can be arranged  by taking
$M=Q^{\dag} Q$ where $Q$ is the Wilson-Dirac fermion matrix. Then $M$
will be sparse (it has at most  next to nearest neighbour  terms) and
the Monte Carlo process can be simulated efficiently. Taking $\alpha =
1,\dots N $ samples of the scalar fields  $\phi^{\alpha}_i$
yields stochastic estimates of the inverse of $M$. Thence the 
propagator $Q^{-1}$ can be recovered from these  samples  since
$G_{ji}=Q^{-1}_{ji}=\, <(Q_{ik} \phi_k)^* \phi_j>$.

  The stochastic evaluation of propagators has the advantage that the 
number of correlations measured increases hugely - by $L^3 T$ for a
lattice  of spatial size $L$ and time length $T$. This is compensated by
the increase in  error from the stochastic (rather than exact)
propagators used. The statistical error is easily estimated since each 
scalar field $\phi_i$ has a variance of order 1 (here we
normalise the  Wilson-Dirac matrix as $Q=1-KD$ with hopping parameter
$K$).  Thus for $N$  samples each propagator will have errors of order
$N^{-1/2}$. This  will result in a statistical error on hadron
correlations which is independent of  time separation $t$. So for small
$t$ one will have small relative errors while at larger $t$ the relative
error  will be huge since the signal is so small.  Since hadron
correlators at  reasonably large $t$ values are needed to separate out
ground states and excited states,  a way to reduce these statistical
errors is essential.

One suggestion~\cite{bermion} is to use a local multi-hit variance
reduction for the $\phi$ fields.  This is equivalent to taking an
improved estimator for the $\phi$ field which  is the average over many
Monte Carlo updates of that field with its neighbours held fixed. This
is very easy to implement and provides a significant error reduction. 
This approach has been used for B-meson physics
successfully~\cite{bermion}. Such a variance reduction is a welcome
amelioration but still does not give an  absolute error decreasing with
$t$. This very attractive goal can, however, be achieved and here we
present our proposals.

Imagine that instead of a multi-hit update of one site in the presence
of its neighbours  held fixed, one considered the  updating of
a fixed region P. Let the  scalar fields within P be labelled
$\phi$ and those on the boundary of P be labelled $s$. Then the required
improved estimate of $\phi$ with $s$ held fixed is given by
 \begin{equation}
  v_i= {1 \over \cal{Z}}\int D \phi \,  \phi_i e^{ -{1 \over 2}( \phi_j^*
 \hat{M}_{jk} \phi_k + \phi_j^* \tilde{M}_{jk} s_k + h.c.)}   
 \end{equation}
 where we have distinguished the elements of $M$ connecting  the set of
$\phi$ fields inside P to themselves ($\hat{M}$) and those connecting 
them to the boundary ($\tilde{M}$). The integral over $\phi$ is gaussian
and we obtain:
 \begin{equation}
   v_i = -\hat{M}^{-1}_{ij} \tilde{M}_{jk} s_k
 \end{equation}
 This can be visualised as  the propagator $\hat{M}^{-1}$ in region P 
from source  $\tilde{M} s$ at the boundary to site $i$. We will call $v$
the maximally  variance reduced stochastic estimator. If site $i$ lies 
a minimum of $d$ links from the nearest boundary point, then we would
expect  $v_i$ to decrease in magnitude as $d$ increases. In the lowest
order  hopping parameter expansion this behaviour will be as $K^d$.
Since the variance of $s$ will remain of order one, this implies that
the variance of  $v$ will be maximally reduced by taking a partition P
which has a boundary as far  as possible from  the point $i$. However,
we must remember that this variance reduction formula  applies  if just
{\em one} $\phi$ field within P is to have its variance reduced. To 
evaluate a variance reduced propagator one must choose two disjoint
regions P and R and solve for the variance reduced fields $v_i$ and
$w_j$ in P and R respectively. Then we will have   $G_{ji}=\ <v_i^*
w_j>$  with no bias. Thus the rules of operation are to take regions P
and R around points $i$ and $j$  such that the boundaries of P and R are
as far from these points as possible.

One geometry which allows this is to take P as the time slices $0 < t <
T/2$ and R as the time slices $T/2 < t < T$. The boundary sources S are
then the time slices at $t=0$ and $T/2$.  Then propagators between  P
and R (also between P and S and between R and S) can be evaluated with
maximal variance reduction. This is not  all-to-all just most-to-most,
although by using in turn several different choices of  partitioning
into P and R, all-to-all can be obtained. We have achieved the goal of
stochastic propagators  from most sites to most sites but at what
computational cost? Each variance  reduction operation corresponds to a
sparse matrix inversion with a  source $\tilde{M} s$ (where $s$ are the
stochastic samples of $\phi$ on the source planes at $t=0$ and $T/2$) in
a region less than half the lattice size. N=12 samples of $v$ and $w$
can be obtained for a computational effort equivalent to  one usual
inversion of $M$ for all 12 colour-spin components at one source. This
increases by a factor of 2 or so the computation of evaluating by Monte
Carlo the stochastic fields in the first place. This is more  than
balanced by the huge reduction in error as we shall see. 

 Since this procedure is conceptually a triply nested Monte Carlo, we
summarise  the steps needed. 
 \begin{itemize}
  \item  Create  gauge configurations $g$.
  \item  In $g$, create  sample scalar fields for all $x$ and
colour-spin components by Monte Carlo using equation~1: $\phi^{\alpha}(g)$.
  \item  Using the scalar field as a source on time-planes at
$t=0$ and $T/2$,  solve iteratively using equation~3 for the improved
estimators $v^{\alpha}(g)$ and $w^{\alpha}(g)$.
  \item then $G_{ji}=\sum_g \sum_{\alpha} (Q_{ik} v_k^{\alpha}(g))^*
w_j^{\alpha}(g)$.
 \end{itemize}

This procedure will certainly provide maximally variance reduced
stochastic  estimators of propagators.  The number of the  stochastic
samples $N$ of the scalar field  should be chosen so that the
resulting error on the quantity of interest from one gauge configuration
is small enough but not  so small that it is less than the intrinsic
variation from one gauge configuration to another.  We explore first the
case of the B meson  using a static approximation for the heavy quark.
This is a very favourable case for stochastic  inversion since there is
a large increase in statistics (through using all  space points as
sources) and the hadronic  correlation  of interest is  proportional to
the light quark propagator so  the choice of $N$ is not critical. Our
choice of partitions P and R is also seen to be  appropriate because 
propagators are only needed in the time direction for which the 
variance reduction will have maximal effect.

\section{B SPECTROSCOPY}

 We consider a small lattice ($8^3$ 16) at  $\beta=5.74$ with Wilson
fermions of hopping parameter $K=0.156$. This choice was motivated by
pre-existing studies~\cite{wupper}. As a first example we  evaluated the
B meson correlator at time separation $t$ using local hadronic 
operators at source and sink. Then we compared conventional inversion
with  various implementations of  stochastic inversion. Sample results
are shown in Table~1 where the comparison has been made  for equal disk
storage of propagators or scalars. We chose $N=25$ samples  to
evaluate the stochastic estimates of the propagators. In the maximally
reduced variance method, $v_i$ and $w_j$  were chosen symmetrically
either side of the source time planes.

\begin{table*}[hbt]
\setlength{\tabcolsep}{1.5pc}
\newlength{\digitwidth} \settowidth{\digitwidth}{\rm 0}
\catcode`?=\active \def?{\kern\digitwidth}
\caption{B meson correlators at $t=7$.}

\begin{tabular*}{\textwidth}{@{}l@{\extracolsep{\fill}}llcc}
\hline
Method & $C(7) \times  10^7$ & Data Set &  CPU \\
\hline
MR inversion  & 3712(147) & propagators from 4 sources  & 1\\
 & & for 10 gauge fields & \\
\hline
Stochastic inversion & & 25 samples of $\phi$ &  \\
\ Basic & 2754(926) & for 20 gauge fields & 2 \\
\ Local multihit & 3418(410) &    &2 \\
\ Maximal variance reduction & 3761(21) &  & 4 \\
\hline

\end{tabular*}
\end{table*}

The maximal variance reduction gives a factor of 7 improvement in error 
compared to other methods for only an overall computational increase of
a factor of 4. This is equivalent to  a net gain of a factor of 12 in
computing time for a similar result. Moreover, the stochastic method
allows  correlations involving different sources (smeared, fuzzed, etc)
to be constructed at little  extra cost. This can be seen in Figure~2
which shows the effective mass  plots for some of these correlators --
these  results come from stochastic evaluation from 20480 sources -- as
described in Table~1. The results~\cite{wupper} from conventional
inversions  (with 170 sources) are seen to be significantly less
precise than those obtained here.

\begin{figure}[t]
\vspace{2.4in}
\includegraphics{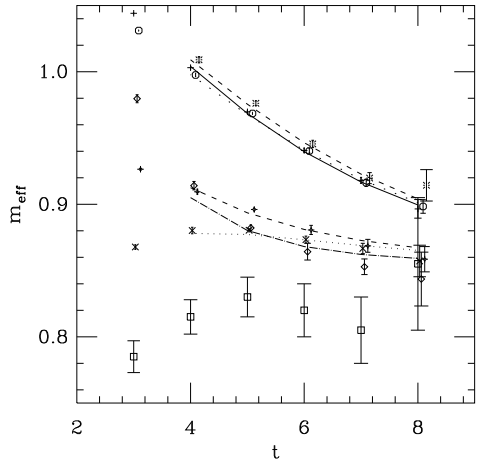}

\caption{  The  B meson effective mass  versus $t$ from our data with
different combinations of  local, fuzzed and $\gamma_i D_i$ sources and
sinks  together with a three exponential fit.  Also shown (squares) are
the Wuppertal data~{\protect\cite{wupper}} for smeared source and local
sink from 170 propagators.}

\end{figure}

Also illustrated is a three state fit to our hadronic correlations which
yields $ma=0.855(6)$ for the ground state and $m'a=1.233(15)$ for the
first excited state of the same quantum numbers,  where the errors
quoted are statistical only. We have also extracted the correlations
corresponding  to the L-excited states ($P$ and $D$-wave). Of particular
interest is the splitting of the $P$ wave level  into two energy values 
in the static heavy quark limit,  corresponding to  $J^{P}= 0^+,\  2^+$
mesons. Our preliminary results are $ma=1.35(5)$ and $1.24(4)$
respectively,  where the errors are purely statistical.   Note that, if
the level ordering indicated by these preliminary results  should be
substantiated, this is a first indication of a meson system for which
the $2^+$  state is lighter than the $0^+$. This is not unexpected in
potential based models  since the relevant spin-orbit splitting has this
sign at large distance~\cite{eichten,cmls}. Further work using larger
lattices, several lattice spacings and  several hopping parameters is
under way to explore this further.

\section{HADRONIC OBSERVABLES}

    For observables with more than one light quark propagator, it is
possible to use the stochastic method,  but some care is needed. One way
to grasp the subtlety is to imagine that  there are two  quarks with 
different flavours  and to split the sum over samples $\alpha=1,\dots N$
into  subsets for  each flavour. Provided these two subsets are
independent, then propagators  of each flavour are obtained without bias.
In practice, provided the stochastic estimators  $\phi^{\alpha}$ are
independent of $\alpha$, then a sum over $\alpha,\ \beta=1,\dots N$ with
$\alpha \ne \beta$ gives the required information. This illustrates that
the  number of samples is effectively $N^2$ for mesons and $N^3$ for
baryons. Indeed  we have observed that the error from the stochastic
sampling reduces as $1/N$ for correlators involving two light quarks in
the region of small $N$. For large $N$ values, there will be no 
advantage in  reducing the error from one gauge sample to below the
intrinsic variation  between gauge samples.
 
    We discuss briefly the potential advantages of using stochastic 
propagators for different hadronic observables. As we have shown above, 
for heavy quarks in the static limit, light quark propagators are needed
in the time direction and these are efficiently produced by the stochastic 
method with maximal variance reduction. Thus other static observables such as  
$\Lambda_b$ and the matrix element $B_B$ are efficiently evaluated.

    For mesons made of light quarks, it is necessary to project the 
correlations into a specific momentum state. This implies a sum of
correlations over  the whole spatial volume at that time. Even though
the signal will be localised at small difference between spatial
coordinates of source and sink, in this sum the  noise will come from
all spatial sites. Another way to see this is that  the propagator for
$t$ time steps and $x$ space steps will behave in the lowest order
hopping parameter expansion as $K^{t+x}$ whereas the variance reduction
we have employed above will  reduce the error by $K^t$ only.  Thus it
appears that stochastic  inversion will require a relatively large
number of samples $N$ to reduce  this noise  in the case of the momentum
zero  hadronic correlation.

 One interesting area of application is to disconnected quark diagrams -
as needed  for $\eta$ meson studies and glueball mixing. The maximal
variance reduction technique  does not seem promising since the variance
reduced average of $\phi_i^2$ will involve $\hat{M}^{-1}_{ii}$ which
is not easily obtained computationally (this is after all the origin of
the problem here anyway). It is feasible to use a more local region P
and this has been  studied~\cite{bereta}.

\section{DISCUSSION}
   
   Particularly for dynamical fermion configurations which are
computationally  expensive to obtain, there is a virtue in measuring the
hadronic properties from a gauge configuration as completely as
possible.  We have presented a method to extract correlators  from all
sources to all sinks to achieve this. Moreover the storage requirements
are  not excessive. For the case of correlators  involving one static
quark, the gain in signal is very significant. For  exploration of
observables involving more than one source and one sink,  such as matrix
elements or meson-meson interactions, then our proposal  is very
promising.

\end{document}